\documentstyle[aps,pra,twocolumn,psfig,floats]{revtex}
\draft
\begin{document}

\title{Extremal equation for optimal completely-positive maps}
\author{Jarom\'{\i}r Fiur\'{a}\v{s}ek}
\address{Department of Optics, Palack\'{y} University, 17. listopadu 50,
77200 Olomouc, Czech Republic}

\maketitle

\begin{abstract}
We derive an extremal equation for optimal completely-positive map which
most closely approximates a given transformation between pure quantum
states. Moreover, we also obtain an upper bound on the maximal mean fidelity
that can be attained by the optimal approximate transformation.
The developed formalism is applied to
universal-NOT gate, quantum cloning machines, quantum entanglers,
and qubit $\vartheta$-shifter.
\end{abstract}


\section{Introduction}

It is well known that certain transformations are forbidden in quantum
theory. For example, it is impossible to exactly clone an unknown
quantum state \cite{Wootters82} or to prepare a qubit orthogonal
to a given unknown qubit \cite{Buzek99}.
These transformations can be implemented only approximately
\cite{Buzek96,Gisin97,Niu98,Bruss98,Bruss98b,Werner98}. A convenient
measure of the quality of such an approximate transformation is the
mean fidelity and the transformation is optimal if it reaches the
maximum possible fidelity.

In quantum theory, the class of allowed transformations consists of
trace-preserving completely-positive maps (CP-maps) $\cal{E}$ between
input and output density matrices. In this paper we derive a generic extremal
equation for the optimal CP-map which approximates a given desired
transformation. This equation may easily be solved numerically
by means of repeated iterations which yields the optimal CP-map.
We also derive an upper bound on the maximum attainable
fidelity. If we find a CP-map which saturates this upper bound then
such a transformation is, by definition, the optimal one.
We illustrate the application of our formalism on universal-NOT  gate,
quantum cloning machine, quantum entanglers, and quantum $\vartheta$-shifter.

We shall consider only the unconditional transformations which
always provide an output. This should be contrasted with conditional
transformations whose output is either accepted or rejected in
dependence on the result of some measurement on the ancilla. Note that
the conditional transformations can achieve higher fidelity than
the unconditional ones, but only at the expense of possibly very
low probability of success.

The paper is organized as follows. In Sec. II we introduce a useful
parametrization of the CP-maps and derive the extremal equation
for the optimal CP-map. We also obtain an upper bound on the maximum
fidelity. In Sec. III we briefly describe how any CP-map can be
implemented as a unitary transformation on  larger Hilbert space.
In Sec. IV we present examples of application of our method.
Finally, Sec. V contains conclusions.

\section{Extremal equation and bound on fidelity}

Suppose we would like to implement a transformation
between pure states  $|\psi_{\rm in}\rangle \in \cal{H}$ and
$|\psi_{\rm out}\rangle \in \cal{K}$
\begin{equation}
|\psi_{\rm in}(\vec{x})\rangle \rightarrow
|\psi_{\rm out}(\vec{x})\rangle.
\label{transformation}
\end{equation}
Here $\vec{x}$ denotes a set of numbers parametrizing
all pure states in Hilbert space of input states $\cal{H}$.
Note that the dimensions of input and output Hilbert spaces
may differ in general, ${\rm dim}{\cal{H}}\neq {\rm dim}{\cal{K}}$.
It may happen that the desired transformation (\ref{transformation})
cannot be implemented exactly because (\ref{transformation})
is not a linear CP-map and we thus want to find a CP-map which most
closely approximates the transformation (\ref{transformation}).

Let us begin by introducing a convenient representation of the
CP-maps. We say that map $\hat{\rho}\rightarrow {\cal{E}}(\hat{\rho})$
is positive if it preserves the positivity of operators $\hat{\rho}$.
The map $\cal{E}$ is completely positive if and only if the extension
${\cal{E}}_{\cal{H}}\otimes \cal{I}_{\cal{H}^\prime}$
is a positive map for any Hilbert space $\cal{H}^\prime$, where $\cal{I}$
is an identity map. Any CP-map can be represented by a positive operator
\cite{DAriano01}. Consider the maximally entangled state
on ${\cal{H}}^{\otimes 2}$,
\begin{equation}
|\varphi\rangle=\sum_{j=1}^{\rm dim \cal{H}} |j\rangle_1 |j\rangle_2
\label{entangled}
\end{equation}
and define the operator
\begin{equation}
\hat{\chi}={\cal{E}}_{\cal{H}}\otimes
\cal{I}_{\cal{H}}(|\varphi\rangle\langle\varphi|).
\label{chi}
\end{equation}
Note that $\hat{\chi}$ acts on a Hilbert space
${\cal{H}}\otimes \cal{K}$.
It is easy to show that the CP-map $\hat{\rho}_{\rm
out}={\cal{E}}(\hat{\rho}_{\rm in})$  can be written as
\begin{equation}
\hat{\rho}_{\rm out}={\rm Tr}_{\cal{H}}[\hat{\chi}\,
\hat{\rho}_{\rm in}^{\rm T}\otimes \hat{1}_{\cal{K}}],
\label{cpmap}
\end{equation}
where T stands for the transposition and $\hat{1}_{\cal{K}}$ denotes
the identity operator on $\cal{K}$.
The requirement that the CP-map $\cal{E}$ should preserve the trace
imposes the following constraint on $\hat{\chi}$,
\begin{equation}
{\rm Tr}_{\cal{K}}[\hat{\chi}]=\hat{1}_{\cal{H}}.
\label{tracecondition}
\end{equation}

We would like to quantify how well the CP map $\hat{\chi}$
approximates  the desired transformation (\ref{transformation}).
To this end we define the mean fidelity as
\begin{equation}
F=\int d\vec{x} \langle \psi_{\rm out}(\vec{x})|
{\cal{E}}\left(|\psi_{\rm in}(\vec{x})\rangle \langle \psi_{\rm in}(\vec{x})|
\right) |\psi_{\rm out}(\vec{x})\rangle,
\label{fidelitydef}
\end{equation}
where  $d \vec{x}$ denotes the proper
measure on space of pure states $|\psi_{\rm in}(\vec{x})\rangle$.

With the help of Eq. (\ref{cpmap}) we may rewrite the expression
(\ref{fidelitydef}) as
\begin{equation}
F={\rm Tr}[\hat{\chi} \hat{R}],
\label{fidelity}
\end{equation}
where the positive operator $\hat{R}$ acting on ${\cal{H}}\otimes{\cal{K}}$
is given by
\begin{equation}
\hat{R}= \int d\vec{x}
\left(
|\psi_{\rm in}(\vec{x})\rangle \langle \psi_{\rm in}(\vec{x})|
\right)^{\rm T}
\otimes
|\psi_{\rm out}(\vec{x})\rangle\langle \psi_{\rm out}(\vec{x})|.
\label{Rdefinition}
\end{equation}

Our task is to find a trace-preserving
CP-map $\hat{\chi}$ which maximizes the fidelity (\ref{fidelity}).
We take into account the constraint (\ref{tracecondition})
by introducing an operator Lagrange multiplier
$\hat{\Lambda}=\hat{\lambda}\otimes\hat{1}_{\cal{K}}$
and we look for the maximum of the functional
\begin{equation}
\tilde{F}[\hat{\chi}]={\rm Tr}[\hat{\chi}\hat{R}]
-{\rm Tr}[\hat{\chi}\hat{\Lambda}].
\label{Ftilde}
\end{equation}
We expand $\hat{\chi}$ in eigenstate basis
\begin{equation}
\hat{\chi}=\sum_{j}r_j |\pi_j\rangle\langle \pi_j|,
\label{chiexpanded}
\end{equation}
and rewrite the functional (\ref{Ftilde}) as
\[
\tilde{F}[\hat{\chi}]=\sum_{j}
r_j \langle \pi_j |\hat{R}-\hat{\Lambda}|\pi_j\rangle.
\]
A variation of $\tilde{F}[\hat{\chi}]$ with respect to
$\langle \pi_j|$ yields the extremal equations,
\begin{equation}
(\hat{R}-\hat{\Lambda}) r_j |\pi_j\rangle=0.
\label{piext}
\end{equation}
We note that recently a similar equation has been derived
for elements of positive operator valued measure representing
an {\em optimal quantum measurement} \cite{Derka98,Massar00}.

We multiply Eq. (\ref{piext}) by $\langle \pi_j|$ and sum over $j$.
After some manipulations we obtain
\begin{equation}
\hat{\chi}=\hat{\Lambda}^{-1}\hat{R}\hat{\chi}.
\label{chiasym}
\end{equation}
Further we take Hermitian conjugate of this formula,
 $\hat{\chi}=\hat{\chi}\hat{R}\hat{\Lambda}^{-1}$,
and insert it back to the right-hand side of Eq. (\ref{chiasym}).
Thus we arrive at a symmetrized extremal equation
\begin{equation}
\hat{\chi}=\hat{\Lambda}^{-1}\hat{R}\hat{\chi}\hat{R}\hat{\Lambda}^{-1}.
\label{chiext}
\end{equation}
The Lagrange multiplier $\hat{\Lambda}=\hat{\lambda}\otimes
\hat{1}_{\cal{K}}$ can be determined from the constraint (\ref{tracecondition})
which provides expression for $\hat{\lambda}$,
\begin{equation}
\hat{\lambda}=\left({\rm Tr}_{\cal K}
[\hat{R}\hat{\chi}\hat{R}]\right)^{1/2}.
\label{lambdaext}
\end{equation}
We fix the square root by postulating that $\hat{\lambda}$ is
positive Hermitian operator.
The system of coupled nonlinear extremal equations (\ref{chiext}) and
(\ref{lambdaext}) can be conveniently solved by means of repeated
iterations, starting from some initial `unbiased' CP-map, for example a
map which transforms every input density matrix to the maximally mixed
state on $\cal{K}$, $\hat{\rho}_{\rm in}\rightarrow
\hat{1}_{\cal{K}}/{\rm dim}\cal{K}$.
Note that the iterations preserve the positivity of $\hat{\chi}$ and
the constraint (\ref{tracecondition}) is exactly satisfied at each
iteration step.

From the formula (\ref{fidelity}) we can obtain an upper bound on
the optimal fidelity $F$,
\begin{equation}
F \leq {\rm Tr}[\hat{\chi}] R_{\rm max}={\rm dim}{\cal{H}} R_{\rm max},
\label{Fbound}
\end{equation}
where $R_{\rm max}$ is the largest eigenvalue of the operator $\hat{R}$.
If we find a CP-map which reaches the upper bound on fidelity
(\ref{Fbound})
then such transformation is, by definition, optimal one. We shall use
this theorem below to prove the optimality of quantum cloning machine
and universal-NOT gate.

\section{Implementation of CP-maps}

Before turning to explicit examples of application we should briefly comment
on the possibility of experimental realization of the CP-map $\hat{\chi}$.
It holds that any trace-preserving CP-map can be accomplished
as a unitary evolution on an extended Hilbert space. Any CP-map can be written in the form of Kraus decomposition \cite{Kraus71}
\begin{equation}
\hat{\rho}_{\rm out}=\sum_{l} \hat{A}_l \hat{\rho}_{\rm in}
\hat{A}_l^\dagger
\label{Kraus}
\end{equation}
and the trace-preservation condition gives
\begin{equation}
\sum_l \hat{A}_l^\dagger \hat{A}_l =\hat{1}_{\cal{H}}.
\label{Atrace}
\end{equation}
Rewritten in terms of the matrix elements
$A_{ki}^{(l)}\equiv\langle k|\hat{A}_l|i\rangle$
this constraint reads
\begin{equation}
\sum_{k,l} A_{ki}^{\ast (l)} A_{kj}^{(l)}=\delta_{ij}.
\label{Atraceelem}
\end{equation}
The number of necessary operators $\hat{A}_l$  is equal to
the number of nonzero eigenvalues of matrix $\hat{\chi}$ which we denote
by $C$. In fact, the operators $\hat{A}_l$ may be associated with
the eigenstates $|\pi_l\rangle$ of the operator $\hat{\chi}$,
\begin{equation}
A_{ki}^{(l)}=\sqrt{r_l}\langle k|\langle i|
\pi_{l}\rangle,
\label{Achi}
\end{equation}
where $|i\rangle \in \cal{H}$ and $|k\rangle \in \cal{K}$ are states in
input and output Hilbert spaces, respectively.

In order to implement $\hat{\chi}$ as a unitary transformation, we must
work in Hilbert space with dimension $D=C{\rm dim}{\cal{K}}$,
where we define the transformation
\begin{equation}
|i\rangle |0\rangle \rightarrow \sum_{k,l} A_{ki}^{(l)} |k\rangle
|l\rangle.
\label{unitary}
\end{equation}
Here $|0\rangle$ denotes an initial state of the ancilla and $|l\rangle$
are orthogonal output states of the ancilla. With the help of the formula (\ref{Atraceelem}) it is easy to check that the
transformation (\ref{unitary}) is indeed unitary.

\section{Applications}

Having established the general formalism, we may proceed to explicit
examples. In all the examples, $\cal{H}$ is a Hilbert space of single
qubit, all pure states in $\cal{H}$ can be visualized as points on the
surface of Bloch sphere parametrized by angles $\vartheta$ and $\phi$,
\begin{equation}
|\psi(\vartheta,\phi)\rangle =\cos\frac{\vartheta}{2}|0\rangle
+e^{i\phi}\sin\frac{\vartheta}{2}|1\rangle,
\label{pure}
\end{equation}
and the proper integral measure is
\begin{equation}
\int d \vec{x} \equiv \frac{1}{4\pi}\int_{0}^{2\pi}\int_0^\pi \sin
\vartheta \, d\vartheta \, d\phi.
\end{equation}

\subsection{Universal NOT gate}

Suppose that  we have $N$ copies of an unknown qubit $|\psi\rangle$ a
and we would like to prepare an inverted qubit
$|\psi_\perp\rangle$, $\langle \psi_\perp |\psi\rangle=0$, and
\begin{equation}
|\psi_\perp\rangle= \sin \frac{\vartheta}{2} |0\rangle-e^{i\phi}\cos
\frac{\vartheta}{2}|1\rangle.
\label{psiperp}
\end{equation}
Let us denote the input state consisting of $N$ identically prepared states
$|\psi\rangle$ as
\begin{equation}
|\vec{\psi}\rangle=|\psi\rangle_1 \ldots |\psi\rangle_N.
\label{psivecdef}
\end{equation}
Since it is impossible to exactly invert an unknown qubit, the mapping
\begin{equation}
|\vec{\psi}\rangle \rightarrow |\psi_{\perp}\rangle,
\label{unot}
\end{equation}
can be carried out only approximately \cite{Buzek99}.
The input Hilbert space is a bosonic subspace of
${\cal{H}}^{\otimes N}$, i.e. a Hilbert space ${\cal{H}}_{+}^{\otimes N}$
spanned by totally symmetric
states of $N$ qubits. Let us denote by $|N,k\rangle$, $k=0,\ldots, N$,
a totally symmetric
state where $k$ qubits are in state $|0\rangle$ and $N-k$ qubits are in
state $|1\rangle$. Thus we can write
\begin{equation}
|\vec{\psi}\rangle=\sum_{k=0}^{N} \sqrt{{N \choose k}}
e^{i(N-k)\phi} \left(\cos\frac{\vartheta}{2}\right)^k
\left(\sin\frac{\vartheta}{2}\right)^{N-k} |N,k\rangle.
\label{psivec}
\end{equation}
We calculate the operator $\hat{R}_{\rm NOT}$ as
\[
\hat{R}_{\rm NOT}=\frac{1}{4\pi}\int_{0}^\pi \sin\vartheta d \vartheta
\int_0^{2\pi}  d\phi \,
(|\vec{\psi}\rangle \langle \vec{\psi}|)^{\rm T} \otimes
|\psi_\perp\rangle \langle \psi_\perp|.
\]
After a straightforward integration we arrive at
\begin{eqnarray}
\hat{R}_{\rm NOT}&=& \sum_{k=0}^N
\frac{N-k+1}{(N+1)(N+2)}|N,k\rangle \langle N,k|\otimes |0\rangle \langle 0|
\nonumber \\
&&+\sum_{k=0}^N
\frac{k+1}{(N+1)(N+2)}|N,k\rangle \langle N,k|\otimes |1\rangle\langle 1|
\nonumber \\
&&-\sum_{k=1}^N
\frac{\sqrt{k(N-k+1)}}{(N+1)(N+2)}|N,k\rangle \langle N,k-1|
\otimes |0\rangle\langle 1|
\nonumber \\
&&-\sum_{k=1}^N
\frac{\sqrt{k(N-k+1)}}{(N+1)(N+2)}|N,k-1\rangle \langle N,k|
\otimes|1\rangle\langle 0|.
\nonumber \\
\label{Rnot}
\end{eqnarray}
The matrix $\hat{R}_{\rm NOT}$ is block diagonal and the calculation of the
eigenvalues of $\hat{R}_{\rm NOT}$ boils down to evaluation of eigenvalues
of $2\times 2$ matrices. The largest eigenvalue reads $R_{\rm max}=1/(N+2)$
and it is $N+2$-fold degenerate. Taking into account that
${\rm dim}{\cal{H}}_{+}^{\otimes N}=N+1$  we get immediately from Eq.
(\ref{Fbound}) the upper bound on the fidelity of U-NOT gate,
\begin{equation}
F_{\rm NOT}\leq \frac{N+1}{N+2}.
\end{equation}
Since the fidelity of U-NOT gate proposed by Bu\v{z}ek {\em et al.} reaches
the bound $(N+1)/(N+2)$ it is optimal \cite{Buzek99}. Note also
that our determination of the upper bound on fidelity of the U-NOT
gate is conceptually similar to the derivation of the fidelity of
the optimal quantum measurement \cite{Derka98,Massar95}.

When we insert the operator (\ref{Rnot}) into extremal Eqs.
(\ref{chiext}), (\ref{lambdaext}),  we may numerically
calculate the optimal CP-map. To be explicit, let us consider
the simplest case of a single input copy, $N=1$.
The input and output Hilbert spaces are now identical
and the corresponding operator $\hat{R}_{\rm NOT}$ can be written as
\begin{equation}
\hat{R}_{\rm NOT}= \frac{1}{3}( |01 \rangle\langle 01|
+|10 \rangle\langle 10|
+|\Phi_{-}\rangle\langle \Phi_{-}|),
\label{Runot}
\end{equation}
where
\begin{equation}
|\Phi_{-}\rangle=\frac{1}{\sqrt{2}}
(|00\rangle-|11\rangle).
\label{Phi}
\end{equation}
Notice that $\hat{R}_{\rm NOT}$  is proportional to the projector
to the subspace spanned by vectors $|01\rangle$, $|10\rangle$, and
$|\Phi_{-}\rangle$. Obviously, the eigenvalues of $\hat{R}_{\rm NOT}$
are $1/3$ and $0$ and $F_{\rm NOT}\leq 2/3$.

Due to the specific structure of the operator $\hat{R}_{\rm NOT}$,
there exist infinitely many solutions to extremal equations
(\ref{chiext}) and (\ref{lambdaext}).
In the numerical calculations, the result depends on the initial choice
of $\hat{\chi}$. In all cases,
we obtain an optimal CP-map which reaches the bound $2/3$. If we start to
iterate from the multiple of identity operator, $\hat{\chi}=\hat{1}/2$,
then after a single iteration step we obtain the universal-NOT
gate whose fidelity is state-independent \cite{Buzek99},
\begin{equation}
\left(
\begin{array}{cc}
\rho_{00} & \rho_{01} \\
\rho_{10} & \rho_{11}
\end{array}
\right) \rightarrow
\frac{1}{3}
\left(
\begin{array}{cc}
2\rho_{11}+\rho_{00} & -\rho_{01} \\
-\rho_{10} & 2\rho_{00}+\rho_{11}
\end{array}
\right).
\end{equation}

\subsection{Quantum cloning machine}

An ideal $1\rightarrow N$ cloning machine would prepare $N$ exact clones
of an unknown qubit $|\psi\rangle$,
\begin{equation}
|\psi\rangle \rightarrow |\vec{\psi}\rangle.
\label{cloning}
\end{equation}
We restrict ourselves to the symmetric quantum cloners which produce $N$
identical approximate copies. In this case the output Hilbert state is
the bosonic space ${\cal{H}}_+^{\otimes N}$ and the ideal
output state $|\vec{\psi}\rangle$ is given by Eq. (\ref{psivec}).
The construction of the operator $\hat{R}_c$ is straightforward,
\[
\hat{R}_{c}=\frac{1}{4\pi}\int_{0}^\pi \sin\vartheta d \vartheta
\int_0^{2\pi}  d\phi \,
(|\psi\rangle \langle \psi|)^{\rm T} \otimes
|\vec{\psi}\rangle \langle \vec{\psi}|,
\]
which yields
\begin{eqnarray}
\hat{R}_c&=& \sum_{k=0}^N
\frac{k+1}{(N+1)(N+2)}|0\rangle \langle 0|\otimes
|N,k\rangle \langle N,k|\nonumber \\
&&+\sum_{k=0}^N
\frac{N-k+1}{(N+1)(N+2)} |1\rangle\langle 1| \otimes
|N,k\rangle \langle N,k|\nonumber \\
&&+\sum_{k=1}^N
\frac{\sqrt{k(N-k+1)}}{(N+1)(N+2)} |0\rangle\langle 1| \otimes
|N,k\rangle \langle N,k-1|
\nonumber \\
&&+\sum_{k=1}^N
\frac{\sqrt{k(N-k+1)}}{(N+1)(N+2)} |1\rangle\langle 0| \otimes
|N,k-1\rangle \langle N,k|.
\nonumber \\
\end{eqnarray}
The largest eigenvalue of the block-diagonal matrix $\hat{R}_c$
reads $1/(N+1)$. Consequently, we have the constraint on maximum cloning
fidelity,
\[
F_c\leq \frac{2}{N+1}.
\]
The symmetric $1\rightarrow N$ cloning machines proposed by
Bu\v{z}ek and Hillery \cite{Buzek96} and Gisin and Massar \cite{Gisin97}
saturate the bound $F_c=2/(N+1)$ and are thus optimal, as shown earlier in
Refs. \cite{Gisin97,Bruss98,Bruss98b,Werner98}.

For $N=2$ we solved numerically the extremal Eqs. (\ref{chiext}) and
(\ref{lambdaext}) by means of repeated iterations and we  obtained a
CP-map which describes the optimal universal quantum cloner designed
by Bu\v{z}ek and Hillery \cite{Buzek96}.
The convergence was very fast and we obtained the desired
CP-map with $10$-digit precision after several tens of iterations.
Since the optimal CP-map is unique the numerical solution
does not depend on the choice of initial CP-map $\chi_0$, provided that all
eigenvalues of $\hat{\chi}_0$  are positive.

\subsection{Quantum entangler}

Recently, Bu\v{z}ek and Hillery \cite{Buzek00} investigated the possibility of
entangling an unknown qubit $|\psi\rangle$ with a qubit in known state
$|0\rangle$. They showed that the following entangling transformation
\begin{equation}
|\psi\rangle \rightarrow  {\cal{N}}(|\psi\rangle|0\rangle
+|0\rangle |\psi\rangle)\equiv |\psi_{\rm out}\rangle,
\label{entangling}
\end{equation}
cannot be carried out exactly. The ideal output state
$|\psi_{\rm out}\rangle$ can be expressed as
\begin{equation}
|\psi_{\rm out}(\vartheta,\phi)\rangle
=\frac{\sqrt{2}\cos(\vartheta/2)|00\rangle+
e^{i\phi}\sin(\vartheta/2)|\Psi_{+}\rangle}
{\sqrt{1+\cos^2(\vartheta/2)}},
\end{equation}
and
\begin{equation}
|\Psi_{+}\rangle=\frac{1}{\sqrt{2}}(|01\rangle+|10\rangle).
\end{equation}
Bu\v{z}ek and Hillery constructed a universal
entangler whose state-independent fidelity is $F\approx 0.946$
\cite{Buzek00}.
Here we show that this entangler is not optimal and we find an
entangling machine which achieves a higher fidelity.

We proceed along the same lines as in the two previous examples and
determine the operator $\hat{R}_{\rm ent}$.
After a straightforward integration we find,
\begin{eqnarray}
\hat{R}_{\rm ent}&=&
(2\ln 2 -1) |0\rangle\langle 0| \otimes |00\rangle\langle 00|
\nonumber \\[1mm]
&&+(3-4\ln 2) |1\rangle\langle 1| \otimes |00\rangle\langle 00|
\nonumber \\[1mm]
&&+(3/2-2\ln 2 ) |0\rangle\langle 0| \otimes |\Psi_+\rangle\langle \Psi_+|
\nonumber \\[1mm]
&&+(4\ln 2-5/2) |1\rangle\langle 1| \otimes |\Psi_+\rangle\langle \Psi_+|
\nonumber \\[1mm]
&&+\sqrt{2}(3/2-2\ln2)|0\rangle\langle 1| \otimes |00\rangle\langle \Psi_+|
\nonumber \\[1mm]
&&+\sqrt{2}(3/2-2\ln 2)|1\rangle\langle 0| \otimes |\Psi_+\rangle\langle 00|.
\end{eqnarray}
Since the largest eigenvalue of $\hat{R}_{\rm ent}$
is $1/2$ we have from (\ref{Fbound})
that $F_{\rm ent}\leq 1$. Although the upper bound $1$ cannot be reached
we can get closer than in Ref. \cite{Buzek00}. When we numerically
solve the extremal equations for optimal CP-map, we find that the
optimal map is a unitary transformation on the space of two qubits,
where the second qubit is initially prepared in the state $|0\rangle$,
\begin{eqnarray}
&&|00\rangle \rightarrow |00\rangle, \qquad
|10\rangle \rightarrow |\Psi_{+}\rangle.
\end{eqnarray}
For an arbitrary input qubit (\ref{pure}) we thus have
\begin{equation}
|\psi\rangle|0\rangle \rightarrow \cos\frac{\vartheta}{2}|00\rangle
+e^{i\phi}\sin\frac{\vartheta}{2}|\Psi_{+}\rangle.
\label{entangleunitary}
\end{equation}
The fidelity of this transformation depends on $\vartheta$,
\begin{equation}
F_{\rm ent}(\vartheta)=
\frac{[\sqrt{2}\cos^2(\vartheta/2)+
\sin^2(\vartheta/2)]^2}{1+\cos^2(\vartheta/2)}.
\label{cloningfidelity}
\end{equation}
We plot the fidelity $F_{\rm ent}(\vartheta)$ in Fig. 1.
The mean fidelity reads
\begin{equation}
F_{\rm ent}=3\sqrt{2}-7/2+(6-4\sqrt{2})\ln(2)\approx 0.9805,
\end{equation}
which is clearly higher than the fidelity $0.946$ obtained in \cite{Buzek00}.
Since $F_{\rm ent}(\vartheta)$ is state dependent, the entangler is
not universal. Nevertheless, for any state $|\psi\rangle$, the fidelity of
the transformation (\ref{entangleunitary}) is higher than the fidelity
of the universal entangler proposed in \cite{Buzek00} and the minimum of
the fidelity (\ref{cloningfidelity}) is
\begin{equation}
F_{\rm min}=4\sqrt{2}(\sqrt{2}-1)^2\approx 0.9706.
\end{equation}
It is easy to see why the non-universal entangler outperforms the
universal one. Since we {\em a-priori} know the second state
$|0\rangle$, this information breaks down the symmetry, and there is no
reason to expect that the universal entangler should be the best one.
Indeed, there exists a preferred basis spanned by $|0\rangle$ and its
orthogonal counterpart $|1\rangle$. Note that the transformation
(\ref{entangleunitary})
realizes ideally the desired transformation (\ref{entangling})
for the basis states $|0\rangle$ and $|1\rangle$.

\begin{figure}[t]
\centerline{\psfig{figure=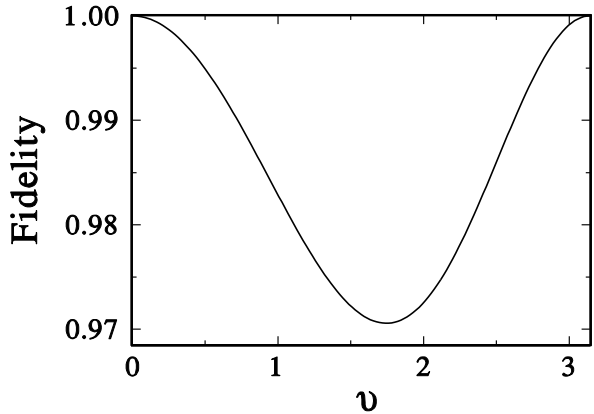,width=0.9\linewidth}}
\caption{Fidelity $F_{\rm ent}(\vartheta)$ of the entangling machine.}
\end{figure}

Let us also study another type of quantum entangler considered in
\cite{Buzek00}.
Suppose we have an unknown qubit $|\psi\rangle$ and we would like to
prepare a maximally entangled state
\begin{equation}
|\psi\rangle \rightarrow
\frac{1}{\sqrt{2}}(|\psi\rangle
|\psi_\perp\rangle+|\psi_\perp\rangle |\psi\rangle).
\label{entangleB}
\end{equation}
On inserting the desired output state (\ref{entangleB}) into
(\ref{Rdefinition}), we find
\begin{equation}
\hat{R}_{\perp}=
\frac{1}{8}\hat{1}_1\otimes(\hat{1}_2\otimes \hat{1}_3+
\frac{1}{3}\sum_{j=x,y,z}\hat{\sigma}_{2,j}\otimes \hat{\sigma}_{3,j}),
\label{RentB}
\end{equation}
where $\hat{\sigma}_j$  denote Pauli matrices and the subscripts $1$
and $2,~3$, refer to input Hilbert space $\cal{H}$ and output space
${\cal{K}}={\cal{H}}\otimes{\cal{H}}$, respectively.
The largest eigenvalue of $\hat{R}_{\perp}$, which is equal to $1/6$,
limits the fidelity of the entangler (\ref{entangleB}) to
\begin{equation}
F_{\perp}\leq 1/3.
\end{equation}
This result was already obtained in \cite{Buzek00}
but our proof is much simpler than that presented in \cite{Buzek00}.
When we iterate  Eqs. (\ref{chiext}) and
(\ref{lambdaext}),
we obtain the optimal CP-map which reaches the maximum fidelity $1/3$.
It turns out that the optimal transformation is strikingly simple.
For any input state, the entangler should prepare the same mixed output state
\begin{equation}
\hat{\rho}_{\rm out}= \frac{1}{3}\left(
|00\rangle \langle 00|+|\Psi_{+}\rangle \langle \Psi_{+}|+
|11\rangle \langle 11|\right).
\label{rhooutperp}
\end{equation}
The fidelity is $1/3$ and does not depend on the input state.
Moreover, it can be easily shown with the help of the Peres-Horodecki
criterion \cite{Peres96,Horodecki96} that the state (\ref{rhooutperp})
is separable.  Thus we conclude that this optimal universal
``entangler'' exhibits rather poor performance because it does not
prepare an entangled state.

\subsection{Optimal $\vartheta$-shifter}

Recently, Hardy and Song \cite{Hardy01,Hardy01b} studied an approximate
implementation of single-qubit transformation where the angle
$\vartheta$ is shifted by $\alpha\in[0,\pi]$,
\begin{equation}
|\psi(\vartheta,\phi)\rangle \rightarrow
|\psi(\vartheta+\alpha,\phi)\rangle.
\label{shifter}
\end{equation}
Hardy and Song found that the optimal universal CP-map
whose fidelity does not depend on $\vartheta$ is identity map for
$\alpha<\pi/2$ and a U-NOT gate for $\alpha >\pi/2$. Remarkably,
the lowest fidelity $F=1/2$ is achieved for $\alpha=\pi/2$.
However, the universal transformation is not the best one and one can
find a CP-map which for certain range of angles $\alpha$
approximates the transformation (\ref{shifter}) much better (i.e.
attains much higher mean fidelity) than the optimal
universal transformation \cite{Hardy01b}.

In what follows we show that our approach naturally and
straightforwardly leads to the optimal non-universal CP-map.
On inserting the output state (\ref{shifter}) into Eq. (\ref{Rdefinition})
we obtain the operator $\hat{R}_\vartheta(\alpha)$,
\begin{eqnarray}
\hat{R}_\vartheta(\alpha)&=&
\left(\frac{1}{4}+\frac{1}{12} \cos\alpha -\frac{\pi}{16} \sin\alpha
\right) |00\rangle \langle 00|, \nonumber \\
&&+ \left(\frac{1}{4}-\frac{1}{12} \cos\alpha +\frac{\pi}{16} \sin\alpha
\right) |01\rangle \langle 01|, \nonumber \\
&&+ \left(\frac{1}{4}-\frac{1}{12} \cos\alpha -\frac{\pi}{16} \sin\alpha
\right) |10\rangle \langle 10|, \nonumber \\
&&+ \left(\frac{1}{4}+\frac{1}{12} \cos\alpha +\frac{\pi}{16} \sin\alpha
\right) |11\rangle \langle 11|, \nonumber \\
&&+\frac{1}{6}\cos\alpha(|00\rangle \langle 11|+|11\rangle \langle 00|).
\end{eqnarray}
We have solved numerically the extremal Eq. (\ref{chiext}),
(\ref{lambdaext}) for various values of $\alpha$.
An analysis of the structure of resulting CP-maps reveals
that they  represent a simple damping process and we can make for them
the following ansatz,
\begin{eqnarray}
|0\rangle \langle 0| &\rightarrow& \cos^2\beta |0\rangle\langle 0|
+\sin^2\beta \,|1\rangle\langle 1|, \nonumber \\[1mm]
|1\rangle \langle 0|&\rightarrow& \cos \beta |1\rangle \langle 0|,
\nonumber \\[1mm]
|0\rangle \langle 1|&\rightarrow& \cos \beta |0\rangle \langle 1|,
\nonumber \\[1mm]
|1\rangle \langle 1|&\rightarrow& |1\rangle \langle 1|.
\label{damping}
\end{eqnarray}
The mean fidelity is a function of $\alpha$ and $\beta$,
\begin{equation}
F(\alpha,\beta)=\frac{1}{2}+\frac{\cos\alpha}{6}( \cos^2\beta+2\cos\beta)
+\frac{\pi}{8}\sin\alpha \sin^2\beta.
\end{equation}
The optimal $\beta$ is obtained from the extremal Eq.
\begin{equation}
\frac{\partial}{\partial \beta}F(\alpha,\beta)=0.
\end{equation}
The solution
\begin{equation}
\beta_1=0
\end{equation}
exists for all $\alpha$. A second solution
\begin{equation}
\cos\beta_2=\left(\frac{3\pi}{4}\tan\alpha-1\right)^{-1}
\label{betatwo}
\end{equation}
exists only if
\begin{equation}
\alpha \geq \arctan \frac{8}{3\pi}=\alpha_0.
\end{equation}
The second root always leads to higher mean fidelity than the first one.
Thus for $\alpha<\alpha_0$ it seems optimal to apply an
identity operation. However, as soon as the angle shift $\alpha$
overcomes the threshold $\alpha_0$ one should rather apply a damping
process (\ref{damping}) with angle $\beta$ given by Eq. (\ref{betatwo}).
The resulting fidelity $F(\alpha)$ can be expressed as
\begin{equation}
F(\alpha)=\left\{
\begin{array}{ll}
\frac{1}{2}(1+\cos\alpha), ~~~ & \alpha\leq\alpha_0,
 \\[1.5mm]
 F(\alpha,\beta_2), & \alpha>\alpha_0,
\end{array}
\right.
\label{Fshifter}
\end{equation}
and is plotted in Fig. 2. Notice that $F(\alpha)$ is not a monotonic
function of $\alpha$ \cite{Hardy01b}.
In Fig. 2 we also show the upper bound on
fidelity obtained from the maximum eigenvalue of
$\hat{R}_\vartheta(\alpha)$.
We can see that the fidelity (\ref{Fshifter}) of the approximate
$\vartheta$-shifter (\ref{damping}) is lower than the upper
bound. This bound is reached only at three discrete points.
The most trivial case is $\alpha=0$ where the identity transformation
achieves fidelity $1$. For $\alpha=\pi$  the U-NOT gate reaches maximum
possible fidelity $2/3$. The third point $\alpha=\pi/2$,
where the fidelity of the $\vartheta$-shifter reads
\[
F\left(\frac{\pi}{2}\right)=\frac{4+\pi}{8},
\]
is perhaps most interesting.
Since for $\alpha=\pi/2$ we have from Eq. (\ref{betatwo}) $\beta_2=\pi/2$,
the output of the $\vartheta$-shifter (\ref{damping}) is the
state $|1\rangle$ regardless of the input state.
This means that the  transformation $\vartheta\rightarrow \vartheta+\pi/2$
is implemented exactly for all pure states on the equator of Bloch sphere
because all these states should be mapped to the south pole of this sphere,
i.e. to the state $|1\rangle$.

\begin{figure}[t]
\centerline{\psfig{figure=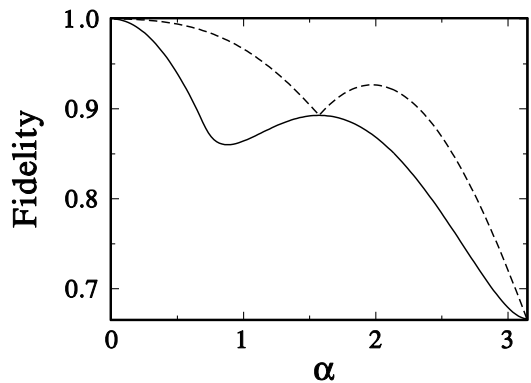,width=0.8\linewidth}}
\vspace*{2mm}
\caption{Fidelity $F(\alpha)$ of the $\vartheta$-shifter
(solid line). The dashed line shows the upper bound on the fidelity
obtained from the largest eigenvalue of $\hat{R}_\vartheta(\alpha)$.}
\end{figure}

\section{Conclusions}

We have derived an extremal equation for optimal completely
positive map $\hat{\chi}$ which most closely approximates certain
prescribed transformation between pure states.
The  nonlinear extremal equation can
be conveniently numerically solved by means of repeated iterations,
which provides very straightforward and simple means for deriving
the optimal CP-map. Moreover, we have obtained an upper bound
on the mean fidelity which, in some cases, allows us to simply
prove the optimality of the calculated CP-map.
We have applied our results
to universal NOT gate, quantum cloning machine, quantum
entanglers, and qubit $\vartheta$-shifter.
The examples discussed in this paper illustrate that the present method is
very general and may find further applications in quantum information
processing, for example in design of optimal CP maps for quantum
teleportation \cite{Rehacek01}.

\acknowledgments
I would like to thank J. \v{R}eh\'{a}\v{c}ek,
Z. Hradil, R. Filip, and M. Je\v{z}ek for
valuable comments and stimulating discussions. This work was supported
by Grant No LN00A015 of the Czech Ministry of Education.

\end{document}